\documentclass[conference, 10pt]{IEEEtran}
\IEEEoverridecommandlockouts
\usepackage{todonotes}
\setuptodonotes{color=blue!30}
\usepackage[T1]{fontenc} 
\usepackage[cmintegrals]{newtxmath}
\usepackage{bm}
\usepackage[numbers, sort&compress]{natbib}
\usepackage{amsmath,amsfonts} 
\usepackage{multirow, tabularx, booktabs}
\newcolumntype{Y}{>{\centering\arraybackslash}X} 
\newcolumntype{Z}{>{\raggedleft\arraybackslash}X} 
\usepackage{algorithmic}
\usepackage{graphicx} 
\graphicspath{ {./figures/} }
\usepackage[caption=false]{subfig}
\usepackage[export]{adjustbox}
\usepackage{textcomp}
\usepackage{pgfplots}
\usepackage{pgfplotstable} 
\usepgfplotslibrary{patchplots, colormaps, groupplots, dateplot}
\pgfplotsset{compat=1.18}
\usetikzlibrary{arrows, chains, positioning,fit,calc, shapes, shapes.geometric, matrix, decorations.pathmorphing, automata, shapes.gates.logic.US,shapes.gates.logic.IEC,calc, external}
\usepackage{xcolor}
\usepackage{soul}
\sethlcolor{yellow}

\definecolor{hsurot}	{cmyk}{.00 1 .59 .26}
\definecolor{hsugrau}	{cmyk}{.38 .37 .39 .15}
\definecolor{hsugelb}	{cmyk}{0 .16 .80 0}
\definecolor{hsublau}	{cmyk}{1 .40 0 .82}
\definecolor{hsuturkis}	{cmyk}{1 .14 .60 .49}
\definecolor{hsugruen}	{cmyk}{.16 .16 .91 .28}
\definecolor{hsubraun}	{cmyk}{.00 .57 1 .17}
\definecolor{hsuorange}	{cmyk}{.01 .87 .77 .13}

\usepackage{listings}
\usepackage[printonlyused, nohyperlinks]{acronym}
\acrodef{kbre}[KBRE]{knowledge-based requirements engineering}
\acrodef{llm}[LLM]{large language model}
\acrodef{msosa}[MSOSA]{CATIA Magic Systems of Systems Architect}
\acrodef{re}[RE]{requirements engineering}
\acrodef{cot}[COT]{chain-of-thought}

\acrodefplural{llm}[LLMs]{large language models}
\usepackage[english]{babel} 	
\usepackage{hyperref}
\hypersetup{hidelinks=true, colorlinks=false, unicode=true, linkcolor=[rgb]{0.10,0.05,0.67}, citecolor=[rgb]{0.10,0.05,0.67}, filecolor=[rgb]{0.10,0.05,0.67}, urlcolor=[rgb]{0.10,0.05,0.67}}
\addto\extrasenglish{}

\addto\extrasenglish{}
\addto\extrasenglish{}
\addto\extrasenglish{}
\addto\extrasenglish{}
\addto\extrasenglish{}
\newcommand{\figspace}{\vspace{-0.175cm}}
\begin{document}
\bstctlcite{IEEEexample:BSTcontrol}
\title{Exploring LLMs for Verifying Technical System Specifications    Against Requirements\\
\thanks{This research in the projects OptiFlex and EKI was funded by dtec.bw – Digitalization and Technology Research Center of the Bundeswehr.  dtec.bw is funded by the European Union – NextGenerationEU.}
}
\author{\IEEEauthorblockN{Lasse M. Reinpold, Marvin Schieseck, Lukas P. Wagner, and Felix Gehlhoff}
\IEEEauthorblockA{\textit{Institute of Automation Technology}\\
Helmut Schmidt University Hamburg, Germany \\
Email:\{lasse.reinpold, marvin.schieseck, lukas.wagner, felix.gehlhoff\}@hsu-hh.de
}
\and
\IEEEauthorblockN{Alexander Fay}
\IEEEauthorblockA{\textit{Chair of Automation} \\
Ruhr University, Bochum, Germany\\
Email: alexander.fay@rub.de}} 
\maketitle

\begin{abstract}
Requirements engineering is a knowledge intensive process and crucial for the success of engineering projects. The field of knowledge-based requirements engineering (KBRE) aims to support engineers by providing knowledge to assist in the elicitation, validation, and management of system requirements. The advent of large language models (LLMs) opens new opportunities in the field of KBRE. This work experimentally investigates the potential of LLMs in requirements verification. Therein, LLMs are provided with a set of requirements and a textual system specification and are prompted to assess which requirements are fulfilled by the system specification. Different experimental variables such as system specification complexity, the number of requirements, and prompting strategies were analyzed. Formal rule-based systems serve as a benchmark to compare LLM performance to. Requirements and system specifications are derived from the smart-grid domain. Results show that advanced LLMs, like GPT-4o and Claude 3.5 Sonnet, achieved f1-scores between 79~\% and 94~\% in identifying non-fulfilled requirements, indicating potential for LLMs to be leveraged for requirements verification.
\end{abstract}

\begin{IEEEkeywords}
Large Language Model, Requirements Verification, Requirements Engineering
\end{IEEEkeywords}
\newcommand*\emptycirc[1][1ex]{\tikz\draw[thick] (0,0) circle (#1);} 
\newcommand*\halfcirc[1][1ex]{%
  \begin{tikzpicture}
  \draw[fill] (0,0)-- (90:#1) arc (90:270:#1) -- cycle ;
  \draw[thick] (0,0) circle (#1);
  \end{tikzpicture}}
\newcommand*\fullcirc[1][1ex]{%
  \begin{tikzpicture}
  \draw[thick, fill] (0,0) circle (#1);
  \end{tikzpicture}}
\newcommand*\notapplicable[1][1.75ex]{\tikz\draw[fill, hsugrau] (0,0) rectangle (#1,#1);}
\section{Introduction} \label{sec:intro}
Requirements engineering (\acsu{re}) is crucial to ensuring that project objectives are clearly defined, understood, and aligned with stakeholder needs, ultimately reducing the risk of errors, delays, and project failure. \ac{re} is also knowledge-intensive and error prone \cite{Maalej.2013c}. The field of \ac{kbre} aims to assist engineers during the \ac{re} process by providing them with domain-specific knowledge, thereby improving accuracy and reducing errors in the development of complex systems \cite{Ratchev.2003}.

Conventionally, \ac{kbre} relies on rule-based or case-based expert systems. Both approaches require the formalization of information, either as structured rules (e.g., SWRL rules, OCL constraints) or as information models documenting past problem-solving experiences. A key challenge of both methods is that they commonly require knowledge engineers to formalize information often originating from texts, and this formalized information is often not intuitively understandable to non-experts \cite{Bashir.2024}.

The emergence of \acp{llm} offers new opportunities in the field of \ac{kbre}: \acp{llm} can process textual information, potentially alleviating the need for formalizing knowledge. Additionally, \acp{llm} comprise a vast amount of pre-trained information, which decreases the reliance on predefined rules to model every piece of knowledge, as is required in rule-based expert systems. This, in turn, can lower the effort required to build and maintain a comprehensive knowledge base. A textual knowledge base offers the potentially added benefit of being more intuitively understandable for users compared to a formalized knowledge base. This can make knowledge-based systems more accessible to non-expert users. However, it remains unclear whether \acp{llm} can consistently generate accurate inferences for \ac{re} tasks, particularly due to potential issues like hallucination \cite{Martino.2023}.

\begin{figure*}[h]
    \centering
    \includegraphics[width=.7\textwidth]{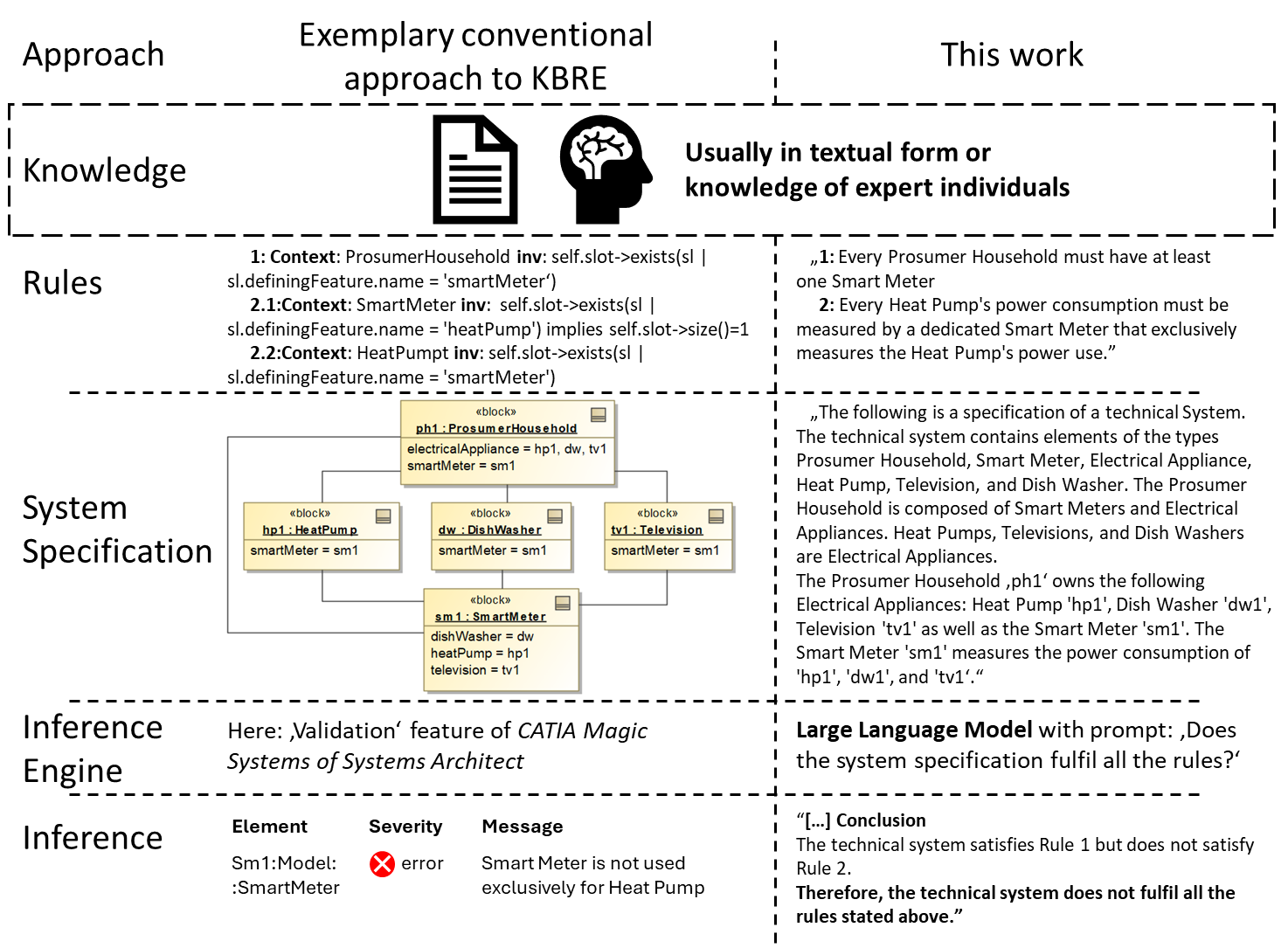}
    \vspace{-.35cm}
    \caption{Comparison of rule-based and \acp{llm}-based approach to evaluate system specifications against requirements.}
    \vspace{-0.5cm}
    \label{fig:LLMvsRule}
\end{figure*}

This work explores the potential of \acp{llm} in performing inference tasks during \ac{re}, by assessing the correctness of inferences made regarding the fulfillment of requirements. Experiments are designed to compare the inferences made by \acp{llm} to rule-based expert systems, which guarantee a correct evaluation and are therefore used as a benchmark. Several scenarios are investigated where an \ac{llm} is provided with a textual system specification and a set of requirements and is prompted to assess which of the requirements are fulfilled by the system specification. The rule-based approach to evaluate system specifications against requirements and an equivalent \ac{llm}-based approach are shown in \autoref{fig:LLMvsRule}. In \autoref{fig:LLMvsRule}, the rule-based approach consists of a SysML model representing the system specification and several OCL constraints representing the requirements. For the \ac{llm}-based approach, the system specification and requirements are formulated in an equivalent form using natural language. 

The experiments are intended to simulate a situation in which an engineer wants to evaluate a system specification in early development stages to assess if the specification fulfills a set of requirements. Such an approach can be especially beneficial if engineers are faced with problems in unfamiliar domains. In this case the requirements serve as a knowledge base for the domain.

Summarizing the above paragraphs, the research question addressed in this study is:

    "\emph{Can \acp{llm} perform inferences comparable to those of conventional rule-based systems in determining which requirements from a given list of requirements are fulfilled by a given system specification?}"

The remainder of this work is structured as follows: \autoref{sec:relatedworks} analyzes related work leveraging \acp{llm} to perform inferences in requirements engineering or related use cases. \autoref{sec:methodology} describes the experiments designed to answer the research question. \autoref{sec:results} presents the experimental results. \autoref{sec:conlusion} presents a discussion of the pursued approach, a conclusion, and an outlook to future works. 
\figspace
\section{Related Works} \label{sec:relatedworks}
This section focuses on studies that leverage \acp{llm} for inferences in requirements engineering. The related works are assessed based on four criteria, and the most distinguishing factors are highlighted for every study:

\begin{enumerate}
    \item Are system specifications evaluated against requirements?
	\item Is the evaluation performed by \acp{llm}?
	\item Are system specifications and requirements written in natural language?
	\item Is the evaluation of the \acp{llm} compared to formal methods for automated inference of requirements fulfillment?
\end{enumerate}

\citet{Gartner.2024} introduce ALICE, a system that combines \acp{llm} and formal logic to detect contradictions in formal engineering requirements. While ALICE incorporates \acp{llm}, an important step of the inference is performed by rule-based logic. 

\citet{Preda.2024} explore the use of \acp{llm} to assess coverage between high-level and low-level requirements in safety-critical systems. The \acp{llm} demonstrate high accuracy in detecting incomplete coverage. The study, however, evaluates dependencies between different levels of requirements rather than comparing system specifications with requirements. 

\citet{Liu.2024} investigate generating code for safety-critical systems using \acp{llm} to ensure compliance with textual requirements. This study focuses on verifying code against requirements, with testing as the verification method, rather than evaluating natural language system specifications.

\citet{Couder.2024} analyze whether code fulfills requirements using \acp{llm}. While the study explores \acp{llm} in requirements verification, it differs by focusing on source code rather than textual system specifications. 

\citet{Fazelnia.2024} propose a method for classifying requirements and detecting defects using fine-tuned language models. While their research focuses on inference tasks related to textual requirements, it does not assess system specifications. 

\citet{Kolthoff.2024} evaluate whether graphical user interfaces, abstracted into a specific syntax, fulfill requirements expressed in user stories using \acp{llm}. These abstractions are not in natural language, and formal verification methods are not used for benchmarking \acp{llm} performance. 

\citet{Hassani.2024} uses \acp{llm} to assess legal compliance between data processing agreements and the general data protection regulation, focusing on legal text rather than system specifications or requirements engineering. No formal methods are used for comparison, and neither requirements nor system specifications are evaluated by \acp{llm}. 

In summary, while various studies investigate \acp{llm} in requirements engineering, none focus on evaluating natural language system specifications against requirements and comparing \acp{llm} performance to formal methods.
\figspace
\section{Design of Experiments} \label{sec:methodology}
The purpose of the experiments is to answer the research question of whether \acp{llm} are capable of determining whether requirements are fulfilled by a given system specification. A simplified example of an experiment as conducted in this study is shown in \autoref{fig:LLMvsRule}. In one prompt, an \ac{llm} is provided with the context of the problem to be solved, a set of requirements, a system specification and an instruction to determine whether the system specification fulfills the requirements and, if not, which requirements are not fulfilled. 
In the example in \autoref{fig:LLMvsRule}, a hypothetical \ac{llm} correctly assesses that the system specification is invalid, as the second requirement is not fulfilled. 

In addition to determining which requirements are not fulfilled, the \ac{llm} is instructed to assess if any requirements are not applicable to a given system specification. An example of a non-applicable requirement would be a requirement referring to attributes of a system element that are not present in the system specification. For the exemplary case of \autoref{fig:LLMvsRule}, a non-applicable requirement would be '\textit{Every battery must have a dedicated smart meter}'. As the household does not own a battery, this requirement is non-applicable. There would be no benefit in \ac{llm} falsely identifying this requirement as not-fulfilled and recommending a user to install a smart meter for a non-existent battery. Therefore, the distinction between applicable and non-applicable requirements needs to be investigated in addition to fulfilled and non-fulfilled requirements. 
The following subsections describe the experimental context, i.e., the domain for which system specifications and requirements are formulated (\autoref{sec:expContext}), the experimental approach of comparing \acp{llm} to rule-based systems and the evaluation metrics of the experiments (\autoref{sec:methExpEval}), and the experimental variables investigated in this study (\autoref{sec:expvars}). 

\subsection{Experimental Context} \label{sec:expContext}
The smart grid domain was chosen as the context for the experiments, with all system specifications and requirements related to this domain. The system specifications analyzed by \acp{llm} describe workflows for optimizing energy consumption schedules of prosumers within a smart grid. These specifications detail the various workflow steps and the interactions between different actors in the grid. As an example, \autoref{lst:simpleworkflow} shows the simplest workflow evaluated by \acp{llm} in this study. These workflows are purely demonstrative and do not reflect current legislation or the state of the art in smart grid technology.

\begin{lstlisting}[showstringspaces=false, label={lst:simpleworkflow}, caption=Simple workflow to be evaluated by \acp{llm}]
The following is a description of a practical implementation of the Smart Grid Optimization Workflow.
Once the workflow is started the Prosumer performs the action Optimize Schedule, and an output object Energy Schedule is created and sent to the Grid Operator as input for the actions Validate Schedule and Log Information.
Upon receiving the Energy Schedule object as input, the Grid Operator performs the action Validate Schedule.
Subsequently, two actions can be triggered: if the Energy Schedule Object is invalid, the Prosumer performs the action Optimize Schedule again. If the Energy Schedule object is valid, the Grid Operator uses the Energy Schedule object as input for the action Log Information. 
The action Log Information is performed by the Grid Operator. 
Subsequently, the workflow terminates.
\end{lstlisting}
\figspace

The requirements define constraints for how a workflow may be structured, e.g., requirements define which actions must be present in the workflow or what type of information is exchanged in the workflow. Examples of requirements for the workflow specification above are shown in Listings~following: 

\begin{lstlisting}[showstringspaces=false, label={lst:requirementsexpl}, caption=Exemplary requirements for system specification in Listing 1]
Req 1: "There must be at least one action Set Optimization Horizon in the workflow." --- not fulfilled
 
Req 2: "Every action Validate Schedule must have at least one input and accepts only inputs of the type Energy Schedule." --- fulfilled
 
Req 3: "Every action Predict Price may only be performed by the actor Prediction Services." --- not applicable  
\end{lstlisting}
\figspace

\subsection{Evaluation of Experiments} \label{sec:methExpEval}
\begin{figure}[h]
    \centering
    \includegraphics[width=.5\textwidth]{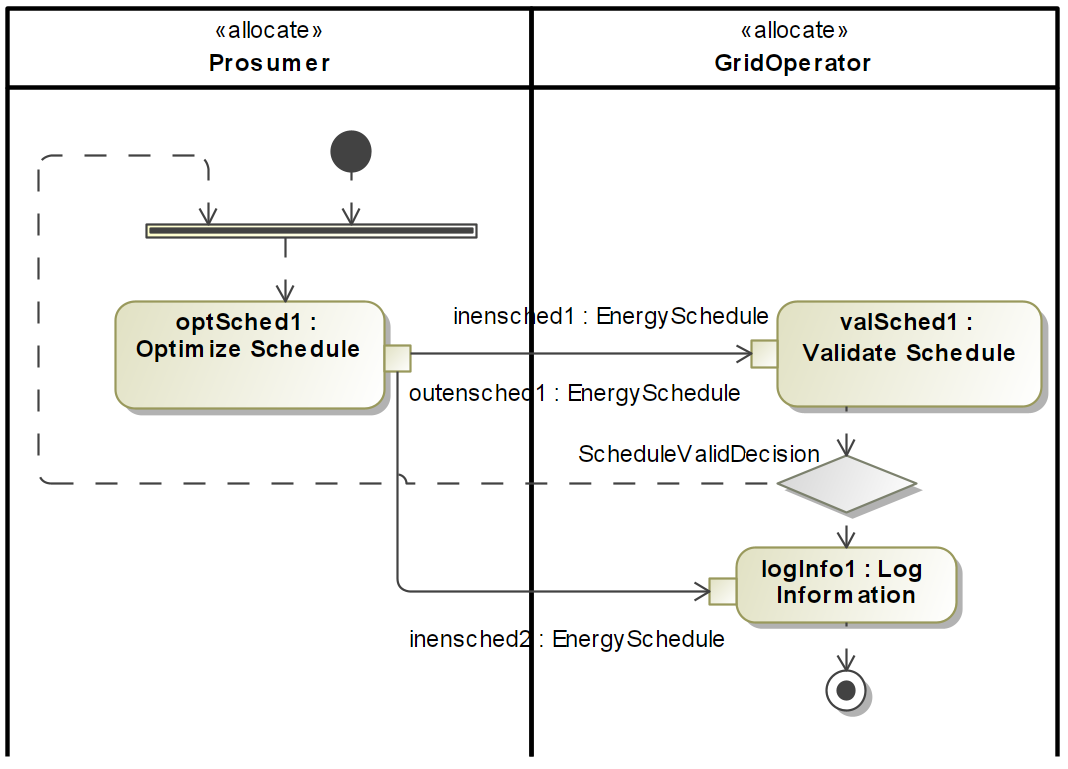}
    \vspace{-.35cm}\vspace{-.35cm}
    \caption{Exemplary smart grid workflow}
    \label{fig:SGworkflow}
\end{figure}

In order to be able to compare the evaluation of system specifications and requirements performed by \acp{llm} to rule-based expert systems, equivalent SysML diagrams of the system specifications are created in \ac{msosa}. Each requirement is formulated as an OCL~constraint. The OCL~constraints are applied to evaluate if the SysML models fulfill the requirements by using the \emph{Validation}-feature of \ac{msosa}. This way, a ground truth is created, which the evaluations of \acp{llm} can be compared to. The smart grid workflow described in the previous section is shown  as a SysML activity diagram in \autoref{fig:SGworkflow}. 

The quality of the evaluation of system specifications by \acp{llm} is measured by precision, recall, and f1-score (see, e.g., \cite{Lipton.2014}. These metrics are calculated both for the evaluation of the fulfillment of requirements by a given system specification and for the evaluation of the applicability of the requirements. This means that for every evaluation of a system specification against a set of requirements, two values for precision, recall, and f1-score are calculated. Therein, a non-fulfilled requirement that was correctly identified as non-fulfilled by the \ac{llm} is a true positive. Likewise, a non-applicable requirement that was correctly identified as non-applicable is also a true positive. 
\figspace

\subsection{Experimental Variables} \label{sec:expvars}
The quality of the evaluation of system specifications by \acp{llm} is evaluated for experimental variables. Each investigated variable is described in the following. 

\textbf{System specification complexity} - The complexity of the workflows described by the system specifications is varied, by altering the number of actors, actions, and objects mentioned in the specification. Three different degrees of complexity are investigated, the least complex consisting of roughly five types of actors, actions, and objects, and the most complex consisting of roughly twenty. 

\textbf{Number of non-fulfilled requirements} - For every degree of system specification complexity, three system specifications exist, one of which fulfills all requirements, one of which does not fulfill one or two requirements, and one of which does not fulfill three of four requirements. In every set of requirements, 20~\% of requirements are non-applicable.

\textbf{Number of requirements} - The number of requirements imposed on every system specification is varied. Every system specification is evaluated against sets of requirements of five, ten, and twenty.

\textbf{Prompting strategy} - Every combination of a system specification and a set of requirements is evaluated by different prompts: a base prompt, instructing the \acp{llm} to evaluate whether the system specification is valid, a \ac{cot} prompt providing the \acp{llm} with detailed instructions on how to structure the evaluation and describe their inferences in detail, and a few-shot prompt featuring one example describing entirely different actions, actors, and objects \cite{Kojima.2022}.

\textbf{LLM} - Four different \acp{llm} are used in the experiments and compared to one another: GPT-4o, GPT-3.5-turbo, Gemini-1.5, and Claude 3.5 Sonnet.

\textbf{Textual style of system specification} - Every system specification is written once in a very structured, clear, and concise style and once in a more conversational style to assess if performance depends on varying styles of potential future users. 

Every combination of the experimental variables described above was tested and repeated three times, leading to a total of 1944 evaluations performed by the \acp{llm}. Experiments were conducted between September 17 and October 7, 2024. All experimental materials including the code, system specifications, requirements, results, SysML models, and OCL constraints are available on Github\footnote{\href{https://github.com/lreinpold/ONCON24_Requirements_Verification_by_LLM}{github.com/lreinpold/ONCON24\_Requirements\_Verification\_by\_LLM}}. 
\figspace
\section{Results} \label{sec:results}
The results of the experimental study are presented individually for each investigated experimental variable. The aggregated results are shown in \autoref{tab:results}, where each horizontal section presents the results for one experimental variable presented in \autoref{sec:expvars}. Every entry in \autoref{tab:results} represents an average of 162 experimental runs, except for the entries in the lowest horizontal section, 'Textual Style', where each entry represents 243 runs. Results are detailed for four different \acp{llm}. For each \ac{llm}, precision, recall, and f1-score are presented for both the analysis of the fulfillment of requirements, and the applicability of requirements. Overall, it is evident that Claude 3.5 Sonnet, and GPT-4o show the most promising results. Therefore, the analysis of the results will focus predominantly on these two models. 

{
\scriptsize
\begin{table*}[!h]
    \centering
    \caption{Results of Evaluation}
    \label{tab:results}
    \begin{tabularx}{\linewidth}{p{1cm}*{24}{|Y}|}
    \toprule
       \multirow{3}{*}{Variable}   & \multicolumn{6}{|c|}{GPT-3.5-turbo} & \multicolumn{6}{|c|}{GPT-4o}& \multicolumn{6}{|c|}{Claude 3.5 Sonnet} &  \multicolumn{6}{|c|}{Gemini 1.5}  \\
       & \multicolumn{3}{|c|}{Ful.} &\multicolumn{3}{|c|}{App.} & \multicolumn{3}{|c|}{Ful.} &\multicolumn{3}{|c|}{App.}  & \multicolumn{3}{|c|}{Ful.} &\multicolumn{3}{|c|}{App.}  & \multicolumn{3}{|c|}{Ful.} &\multicolumn{3}{|c|}{App.}     \\
      & P & R & F1 & P & R & F1 & P & R & F1 & P & R & F1 & P & R & F1 &  P & R & F1 &  P & R & F1 &  P & R & F1 \\
\midrule
\multicolumn{25}{l}{\textbf{Specification Complexity}}      \\
Low &\!\!0.45&\!\!0.37&\!\!0.41&\!\!0.82&\!\!0.36&\!\!0.5&\!\!0.76&\!\!0.87&\!\!0.81&\!\!0.99&\!\!0.8&\!\!0.89&\!\!0.68&\!\!0.84&\!\!0.75&\!\!1&\!\!0.75&\!\!0.86&\!\!0.51&\!\!0.65&\!\!0.57&\!\!0.97&\!\!0.58&\!\!0.73\\
Medium&\!\!0.33&\!\!0.23&\!\!0.27&\!\!0.82&\!\!0.3&\!\!0.44&\!\!0.85&\!\!0.88&\!\!0.87&\!\!0.98&\!\!0.87&\!\!0.92&\!\!0.79&\!\!0.84&\!\!0.82&\!\!1&\!\!0.79&\!\!0.88&\!\!0.59&\!\!0.77&\!\!0.67&\!\!0.97&\!\!0.61&\!\!0.74\\
High&\!\!0.45&\!\!0.11&\!\!0.17&\!\!0.92&\!\!0.47&\!\!0.62&\!\!0.93&\!\!0.81&\!\!0.87&\!\!1&\!\!0.94&\!\!0.97&\!\!0.87&\!\!0.77&\!\!0.82&\!\!0.98&\!\!0.86&\!\!0.92&\!\!0.46&\!\!0.56&\!\!0.51&\!\!0.97&\!\!0.51&\!\!0.67\\
 \midrule
\multicolumn{25}{l}{\textbf{Non-fulfilled Requirements}}      \\
None &\!\!n.A.&\!\!n.A.&\!\!n.A.&\!\!0.95&\!\!0.42&\!\!0.58&\!\!n.A.&\!\!n.A.&\!\!n.A.&\!\!0.99&\!\!0.87&\!\!0.93&\!\!n.A.&\!\!n.A.&\!\!n.A.&\!\!0.99&\!\!0.81&\!\!0.89&\!\!n.A.&\!\!n.A.&\!\!n.A.&\!\!0.97&\!\!0.6&\!\!0.74\\
1 - 2&\!\!0.4&\!\!0.27&\!\!0.32&\!\!0.79&\!\!0.34&\!\!0.48&\!\!0.85&\!\!0.74&\!\!0.79&\!\!0.99&\!\!0.87&\!\!0.93&\!\!0.76&\!\!0.73&\!\!0.74&\!\!1&\!\!0.76&\!\!0.87&\!\!0.48&\!\!0.52&\!\!0.5&\!\!0.98&\!\!0.54&\!\!0.7\\
3 - 4&\!\!0.55&\!\!0.22&\!\!0.32&\!\!0.84&\!\!0.37&\!\!0.51&\!\!0.91&\!\!0.92&\!\!0.92&\!\!0.99&\!\!0.87&\!\!0.93&\!\!0.87&\!\!0.87&\!\!0.87&\!\!0.99&\!\!0.83&\!\!0.9&\!\!0.72&\!\!0.74&\!\!0.73&\!\!0.96&\!\!0.56&\!\!0.71\\
\midrule
\multicolumn{25}{l}{\textbf{Number of Requirements}}      \\
5&\!\!0.76&\!\!0.29&\!\!0.41&\!\!0.94&\!\!0.42&\!\!0.58&\!\!0.93&\!\!0.91&\!\!0.92&\!\!1&\!\!0.84&\!\!0.91&\!\!0.94&\!\!0.94&\!\!0.94&\!\!1&\!\!0.83&\!\!0.91&\!\!0.79&\!\!0.74&\!\!0.76&\!\!0.94&\!\!0.51&\!\!0.66\\
10&\!\!0.35&\!\!0.21&\!\!0.26&\!\!0.97&\!\!0.35&\!\!0.52&\!\!0.81&\!\!0.85&\!\!0.83&\!\!1&\!\!0.85&\!\!0.92&\!\!0.79&\!\!0.77&\!\!0.78&\!\!1&\!\!0.81&\!\!0.9&\!\!0.5&\!\!0.63&\!\!0.55&\!\!1&\!\!0.55&\!\!0.71\\
20&\!\!0.3&\!\!0.23&\!\!0.26&\!\!0.8&\!\!0.38&\!\!0.51&\!\!0.79&\!\!0.82&\!\!0.81&\!\!0.99&\!\!0.89&\!\!0.94&\!\!0.64&\!\!0.77&\!\!0.7&\!\!0.99&\!\!0.78&\!\!0.88&\!\!0.41&\!\!0.61&\!\!0.49&\!\!0.96&\!\!0.59&\!\!0.73\\
\midrule
\multicolumn{25}{l}{\textbf{Prompt}}      \\
base&\!\!0.33&\!\!0.04&\!\!0.07&\!\!1&\!\!0.02&\!\!0.04&\!\!0.7&\!\!0.84&\!\!0.76&\!\!0.99&\!\!0.69&\!\!0.81&\!\!0.53&\!\!0.68&\!\!0.6&\!\!0.98&\!\!0.52&\!\!0.68&\!\!0.22&\!\!0.41&\!\!0.29&\!\!0.75&\!\!0.04&\!\!0.08\\
cot&\!\!0.28&\!\!0.26&\!\!0.27&\!\!0.91&\!\!0.25&\!\!0.4&\!\!0.92&\!\!0.85&\!\!0.88&\!\!0.99&\!\!0.93&\!\!0.96&\!\!0.92&\!\!0.88&\!\!0.9&\!\!1&\!\!0.9&\!\!0.95&\!\!0.77&\!\!0.74&\!\!0.75&\!\!0.96&\!\!0.75&\!\!0.84\\
few-shot&\!\!0.57&\!\!0.42&\!\!0.48&\!\!0.84&\!\!0.86&\!\!0.85&\!\!0.93&\!\!0.87&\!\!0.9&\!\!0.99&\!\!0.99&\!\!0.99&\!\!0.93&\!\!0.89&\!\!0.91&\!\!1&\!\!0.98&\!\!0.99&\!\!0.83&\!\!0.83&\!\!0.83&\!\!0.99&\!\!0.91&\!\!0.95\\
\midrule
\multicolumn{25}{l}{\textbf{Text style}}      \\
struct.&\!\!0.5&\!\!0.25&\!\!0.33&\!\!0.89&\!\!0.43&\!\!0.58&\!\!0.91&\!\!0.84&\!\!0.87&\!\!0.99&\!\!0.91&\!\!0.95&\!\!0.85&\!\!0.85&\!\!0.85&\!\!0.99&\!\!0.88&\!\!0.93&\!\!0.56&\!\!0.73&\!\!0.64&\!\!0.95&\!\!0.59&\!\!0.72\\
conv.&\!\!0.33&\!\!0.23&\!\!0.27&\!\!0.82&\!\!0.32&\!\!0.46&\!\!0.78&\!\!0.87&\!\!0.83&\!\!0.99&\!\!0.83&\!\!0.9&\!\!0.7&\!\!0.79&\!\!0.74&\!\!1&\!\!0.71&\!\!0.83&\!\!0.48&\!\!0.58&\!\!0.52&\!\!0.98&\!\!0.55&\!\!0.7\\
\bottomrule
    \end{tabularx}
        \begin{center}
            Ful. - Fulfillment of Requirements. App. - Applicability of Requirements. P - Precision. R - Recall. F1 - f1-Score
    \end{center}
     \vspace{-.75cm}
\end{table*}
}

The analysis of the impact of system \textbf{specification complexity} shows an interesting trend, in that the precision in identifying non-fulfilled requirements increases at higher system complexities, while the recall decreases for both GPT-4o and Claude~3.5. This means that there are fewer cases where \acp{llm} falsely classify requirements as unfulfilled if the complexity of system specifications is higher. However, the recall decreases at higher specification complexities, meaning that non-fulfilled requirements are more frequently falsely classified as fulfilled at higher complexities. Overall, the f1-score for identifying both non-fulfilled and non-applicable requirements is higher at greater complexity levels for GPT-4o and Claude 3.5, indicating that the evaluation quality of these two models improves with increasing complexity. Potential reasons for this need to be investigated, since this trend cannot be observed for Gemini~1.5 and GPT-3.5-turbo. 

The impact of the number of \textbf{non-fulfilled requirements} shows a similar trend as the previously discussed specification complexity. Both precision and recall show a positive trend, for identifying non-fulfilled requirements, when there are more non-fulfilled requirements to be detected in a system specification. The analysis of the applicability of requirements shows no clear trend.

A clear trend can be observed across all models regarding the impact of the \textbf{number of requirements} analyzed in a single prompt. As the number of requirements increases, both precision and recall decrease when identifying non-fulfilled requirements. This decline is expected, as \acp{llm} generally perform better when larger tasks are decomposed into smaller sub-tasks to be solved in separate prompts \cite{Khot.2022}. In this case, such a decomposition could be realized by having the \ac{llm} analyze twenty requirements in several separate prompts instead of one prompt. A similar trend is not observed for the analysis of non-applicable requirements. 

The \textbf{prompt} also has a strong impact on the quality of the results. For all models, and for both, the identification of non-fulfilled and non-applicable requirements, \ac{cot} prompts perform better than the base prompt (except for the precision score of GPT-3.5-turbo in identifying non-fulfilled requirements), and few-shot prompts perform better than \ac{cot} prompts. Therein, it can be observed that the additional benefits of few-shot prompts over \ac{cot} prompts are especially pronounced in Gemini 1.5 and GPT-3.5-turbo. In GPT-4o and Claude 3.5, few-shot prompts mainly improve the identification of non-applicable requirements, with the two models performing almost perfectly (f1-score of 99~\%) in identifying non-applicable requirements when utilizing few-shot prompts. However, few-shot prompts also slightly improve the identification of non-fulfilled requirement in GPT-4o and Claude 3.5. It should be noted that the low recall in identifying non-fulfilled requirements is likely due to the base prompt not clearly instructing the model to highlight non-applicable requirements.

The \textbf{textual style of the system specification} also shows a clear impact on the results. All models perform worse if the system specifications are written in a more conversational style as compared to a more structured style. This indicates that it is important to utilize concise language as well as clear and consistent terminology when utilizing \acp{llm} for evaluation of requirements fulfillment.

Overall, the best performing model in this study is GPT-4o, clearly outperforming GPT-3.5-turbo and Gemini 1.5, and performing equally or slightly better than Claude 3.5 Sonnet in all but three cases, when comparing f1-scores. However, even for the best performing model GPT-4o, f1-scores between 79\% and 92\% are observed, proving clear deficits compared to formal methods. GPT-3.5-turbo is the worst performing model, exhibiting f1-scores of no more than 50\%, indicating that GPT-3.5-turbo should not be used for evaluation of requirements fulfillment, if utilizing more advanced models is financially feasible. The worse performance of Gemini 1.5 compared to GPT-4o and Claude 3.5 Sonnet can in part be attributed to Gemini 1.5 showing a higher likeliness of stating uncertainty in answers, e.g., Gemini 1.5 is more inclined to state that more information is required to assess the fulfillment of a requirement. This can be a desirable feature, especially in cases where there is no single correct answer. In this study however, where sufficient information was available in all cases, it contributes to a deficit in performance compared to GPT-4o and Claude 3.5.  
\figspace
\section{Discussion and Conclusion} \label{sec:conlusion}
This section is divided into recommendations on how to best utilize \acp{llm} for verifying requirements fulfillment, a discussion of the limitations of this study, and an outlook to future research opportunities in the field. 
\figspace
\subsection{Recommendations}
At the time of writing, the most promising \ac{llm} out of the evaluated \acp{llm} is GPT-4o. It should be noted though that GPT-o1, which is an \ac{llm} specifically designed for performing inferences, could not be tested because access is restricted to a limited number users. However, even the best-performing \acp{llm} achieve F1-scores slightly above 90~\% in the best cases of evaluating requirements fulfillment, thus falling short of matching the performance of the formal approach, which served as a benchmark in this study. \ac{llm} performance has been compared to the formal approach only regarding the accuracy of evaluations. Other factors such as the time required to make inferences and the operational costs, where a formal approach would likely perform better, are beyond the scope of this study.

Within this study, no limitations regarding the investigated system specification complexity were observed, indicating that advanced \acp{llm} can be leveraged to evaluate systems up to a complexity of 20 different types of actors, actions, and objects. While it is likely that \acp{llm} are capable of evaluating even more complex systems without severe degradation in performance, no such systems were evaluated within this study. The experimental data further shows that the number of non-fulfilled requirements in a system specification has a slight positive effect on system performance, indicating that \acp{llm} are especially suitable to evaluate system specifications in early stages when more deficits in the specification can be expected, and the cost of fixing errors is still relatively low. Another important conclusion is that, to evaluate a given set of requirements, the number of requirements to be evaluated in a single prompt should be kept low. Instead, several prompts can be used in parallel to evaluate the same set of requirement, to boost the \acp{llm}' performance. The analysis of the impact of different prompting approaches shows that \ac{cot} prompts with clear instructions clearly outperform base prompts with the simple instruction to evaluate the fulfillment of requirements. Performance can be further increased by applying few-shot prompts, if this is feasible for the respective use case. It should be noted that both few-shot prompting and parallel evaluation of requirements in separate prompts would incur higher financial and/or computational cost. The analysis of the impact of the textual style shows that system specification text that is written in a structured and concise style leads to higher accuracy compared to more conversationally written system specifications. This can cause a drop in \ac{llm} performance when interacting with users that are not familiar with established terminology in a certain domain. 

Beyond the investigated experimental variables, that can be altered to improve \ac{llm} performance, many approaches exist to mitigate factually incorrect evaluations by \acp{llm} that could also be investigated in the context of requirements verification by \acp{llm} \cite{Tonmoy.2024}.

\subsection{Limitations of the study}
The main limitations of this study lie in the investigated dataset. No dataset was found containing requirements and system specifications that are suitable for a comparison with a formal, rule-based approach. A new dataset had to be created for this study. Even though the experiments include a large sample size of 1944 prompts and corresponding evaluations, the number of different combinations of system specifications and sets of requirements was relatively small at 27. The reason for this is the substantial manual effort in creating the formal ground truth requiring the formal modelling of the system specification as SysML diagrams, the creation of OCL~constraints, the creation of validation suites, the validation of the SysML models, and the post-processing of formal validation results. Even though this study shows a clear trend of how to apply \acp{llm} to improve the accuracy of the evaluations, it would be beneficial to perform more experiments, e.g., on different types of SysML diagrams, for more complex inference tasks, and in other domains than the smart grid domain, to figure out the boundaries of \ac{llm} capabilities for the evaluation of requirements fulfillment. 

\subsection{Future Works}
The experimental results clearly show that not all requirements can be correctly identified as fulfilled or not fulfilled by \acp{llm}. Apart from identifying false positive and false negative assessments, no distinction was made regarding different types of errors. Identifying common types of errors might open up opportunities to improve LLM performance. To facilitate such research, all experimental results are publicly available on Github. Future research should also focus on investigating if other benefits that \acp{llm} likely provide, such as utilizing pre-trained information and the relative simplicity in creating textual knowledge bases, outweigh the drawback of occasionally making incorrect inferences. As the majority of requirements is correctly evaluated by advanced \acp{llm}, it should also be investigated if incorrect inferences are the main source of error, or if other sources of error, such as potential misunderstandings on the user side, have a stronger impact on the overall perceived usefulness of leveraging \acp{llm} to assist in requirements engineering. To further decrease the repercussions of incorrect inferences, strategies should be investigated that empower users, even those that are relatively unfamiliar with a given domain, to critically evaluate inferences performed by \acp{llm}.
 
\section{Remarks}
\label{sec:conlusion}
This work has been submitted to the IEEE for possible publication. Copyright may be transferred without notice, after which this version may no longer be accessible.

\figspace
\bibliographystyle{IEEEtranN}
\footnotesize{\bibliography{oncon2023_dp}}
\end{document}